\documentclass[aps,superscriptaddress,twocolumn,showpacs]{revtex4}
\usepackage{amsmath}
\usepackage{graphicx}
\begin{document}
\title{Resonant Compton scattering of electromagnetic waves in a quantum plasma}
\author{Bengt Eliasson}
\affiliation{Institute for Theoretical Physics,
Faculty of Physics and Astronomy,
Ruhr University Bochum, D-44780 Bochum, Germany}
\author{Chuan S. Liu}
\affiliation{Department of Physics, University of Maryland, College Park, MD 20742,
USA}

\date{30 May 2013}
\begin{abstract}
We consider the resonant scattering of coherent electromagnetic waves by a Raman-like process in the gamma ray range off electrostatic modes in a quantum plasma using a collective Klein-Gordon-Maxwell model. The growth rates for the most unstable modes are calculated theoretically, and the results are found to be more efficient than incoherent Compton scattering off individual electrons above a critical amplitude of the electromagnetic wave.
The model does not predict Raman scattering off pair modes that exist in the Klein-Gordon-Maxwell model.
The results are relevant for coherent gamma rays created in forthcoming laboratory experiments
or in astrophysical objects.
\end{abstract}

\maketitle

Laser light in the x-ray range can be used to probe collective effects in
warm dense matter \cite{Glenzer,Glenzer09,Neumeyer10} and on
atomic and molecule levels \cite{Hand09}.
Sub-{\AA}ngstr\"om wavelengths have been obtained using x-ray free electron
lasers \cite{Tanaka12}.
On these length scales,
quantum effects play an important role in the dynamics of the electrons.
Both relativistic and quantum effects are thought to come into play in
x-ray free electron lasers \cite{Schroeder01}.
When the wavelength is comparable to or smaller than the Compton length,
relativistic effects have to be taken into account,
and leads to the Compton scattering, pair production and other effects in the
interaction with matter \cite{Knoll00}.
Gamma ray lasers have been proposed theoretically \cite{Baldwin97,Rivlin07,Rivlin10,Tkalya11,Son12}
but has still to be realized in experiments. The production of coherent
gamma radiation would open up the possibilities to explore matter on sub-atomic scales.
Nonlinear interactions of large-amplitude
electromagnetic (EM) waves with the plasma can lead to
parametric instabilities \cite{Drake74} and relativistic effects
\cite{Shukla86,McKinstrie92,Sakharov94,Guerin95,Adam00}. Collective
Klein-Gordon \cite{Takabayasi53} and Dirac equations \cite{Takabayasi56,Takabayasi57} have been used to model
the relativistic interactions between quantum particles and electromagnetic fields
and to derive quantum fluid models for the particles.
Similar models have been employed to study quantum effects in the in the
nonlinear propagation and scattering instabilities in the interaction between
matter and light using Klein-Gordon \cite{Kuzelev11,Eliasson11a} and Dirac
\cite{Eliasson11b} equations for the electrons,
coupled with Maxwell's equations for the electromagnetic field.

The purpose is here to study the a Raman-like process in which a large amplitude electromagnetic wave
in the gamma-ray regime decays into one electromagnetic daughter wave and one electrostatic plasma wave in a quantum plasma. For this purpose, we use a collective Klein-Gordon-Maxwell model \cite{Takabayasi53,Eliasson11a}, where the wave function represents an ensemble of electrons
and where the electron densities and currents enter into the Maxwell equations for
the electromagnetic fields. As a starting point, we use the nonlinear dispersion relation
for a raman-like 3-wave interaction involving a circularly polarized electromagnetic wave
${\bf A}=(1/2){\bf A}_0\exp(i{\bf k}_0\cdot{\bf r}-i\omega_0 t)+$complex conjugate, decaying into
an electromagnetic daughter wave and plasma oscillations. Considering only the
resonant backscattering scenario in Eq. (38) of Ref. \cite{Eliasson11a} with
$\omega=\omega_-$ and ${\bf k}={\bf k}_-$ but neglecting non-resonant
interactions (terms including $\omega_+$ and ${\bf k}_+$), we have
\begin{equation}
  D_L(\Omega,{\bf K})D_A(\omega,{\bf k})= \frac{D_A(\Omega,{\bf K})\omega_{pe}^2 e^2 |{\bf A}_0\times {\bf k}|^2}{4\gamma_A^3 m_e^2 c^2k^2},
\end{equation}
where
\begin{equation}
  D_A(\omega,{\bf k})=c^2 k^2 -\omega^2+\frac{\omega_{pe}^2}{\gamma_A}
\end{equation}
and
\begin{equation}
  D_L(\Omega,{\bf K})=\frac{\omega_{pe}^2}{\gamma_A}-\Omega^2-\frac{(\Omega^2-c^2 K^2)}{4\omega_C^2}D_A(\Omega,{\bf K}),
\end{equation}
describe the electromagnetic daughter wave and electrostatic plasma wave, respectively.
Here $\omega_{pe}=(n_0 e^2/\varepsilon_0 m_e)^{1/2}$ is the electron plasma frequency and
$\omega_C=\gamma_A m_e c^2/\hbar$ is the relativistically corrected Compton frequency,
where $\gamma_A=\sqrt{1+a_0^2}$ is the relativistic gamma factor due to the large amplitude
EM wave, and $a_0=e A_0/m_e c$ is the normalized vector potential of the EM wave.
Furthermore, $n_0$ is the equilibrium electron number density, $e$ the magnitude
of the electron charge, $m_e$ the electron rest mass, $\varepsilon_0$ the electric constant,
$c$ the speed of light in vacuum, and $\hbar$ Planck's constant divided by $2\pi$.
The frequencies and wave vectors are related through the matching conditions
\begin{equation}
  \omega_0=\omega+\Omega
\end{equation}
and
\begin{equation}
  {\bf k}_0={\bf k}+{\bf K},
\end{equation}
where ($\omega_0$, ${\bf k}_0$), ($\omega$,${\bf k}$), and ($\Omega$, ${\bf K}$) is the frequency and wavenumber of, respectively, the pump wave, the EM daughter wave and plasma wave.
The pump is governed by the dispersion relation $D_A(\omega_0,{\bf k}_0)=0$, while for the EM daughter wave and plasma wave, we have $D_A(\omega,{\bf k})\approx 0$ and $D_L(\Omega,{\bf K})\approx 0$.

In order to find approximate solutions of Eq. (1) for the instability,
we set $\Omega=\widetilde{\Omega}+i\Gamma$ and ${\bf K}=\widetilde{\bf K}$,
where $\widetilde{\Omega}$ and $\widetilde{\bf K}$ simultaneously solve $D_L(\widetilde{\Omega},\widetilde{\bf K})=0$ and $D_A(\omega_0-\widetilde{\Omega},{\bf k}_0-\widetilde{\bf K})=0$, and where $\Gamma$ is the growth-rate.
We then have
\begin{equation}
\begin{split}
  &D_A(\omega,{\bf k})=2i\Gamma(\omega_0-\widetilde{\Omega})+\Gamma^2
  \approx 2i\Gamma(\omega_0-\widetilde{\Omega})
\end{split}
\end{equation}
and
\begin{equation}
\begin{split}
  &D_L(\Omega,{\bf K})=\frac{1}{4\omega_C^2}(2i\Gamma\widetilde{\Omega}-\Gamma^2)
  \\
  &\times\bigg(2\widetilde{\Omega}^2-2c^2\widetilde{K}^2-\frac{\omega_{pe}^2}{\gamma_A}
  +2 i \Gamma\widetilde{\Omega}-\Gamma^2-4\omega_C^2\bigg)
  \\
  &\approx \frac{i\Gamma\widetilde{\Omega}}{2\omega_C^2}\bigg(2\widetilde{\Omega}^2-2c^2\widetilde{K}^2-\frac{\omega_{pe}^2}{\gamma_A}-4\omega_C^2\bigg).
  \end{split}
\end{equation}

To solve approximately the resonance conditions $D_A(\omega_0,{\bf k}_0)=0$, $D_A(\omega,{\bf k})=0$ and $D_L(\widetilde{\Omega},\widetilde{\bf K})=0$ using $\widetilde{\Omega}=\omega_0-\omega$ and $\widetilde{\bf K}={\bf k}_0-{\bf k}$,
we assume that all frequencies are much larger than $\omega_{pe}/\gamma_A$, i.e. that
the plasma is strongly under-dense. We then have
\begin{equation}
  \omega_0=c k_0,
\end{equation}
\begin{equation}
  \omega=ck,
\end{equation}
and
\begin{equation}
  \widetilde{\Omega}=-\omega_C+\sqrt{\omega_C^2+c^2 \widetilde{K}^2}.
\end{equation}
Eliminating $\omega_0$, $\omega$, $\widetilde{\Omega}$, and $\widetilde{\bf K}$ using Eqs. (4), (5) and (8)--(10), we obtain the matching condition
\begin{equation}
  c(k_0-k)=-\omega_C+\sqrt{\omega_C^2+c^2 ({\bf k}_0-{\bf k})^2}
\end{equation}
for the scattering of a relativistically strong electromagnetic wave off electrons.
To put the resonance condition in explicit form, we choose a coordinate system such that ${\bf k}_0=k_0\widehat{\bf z}$,
$k_z=k \cos\theta$ and ${\bf k}_\perp=k_x \widehat{\bf x}+k_y \widehat{\bf y}$, $k_\perp=k\sin\theta$,
and $\theta$ is the angle between ${\bf k}$ and ${\bf k}_0=k_0\widehat{\bf z}$ ($\theta>\pi/2$ corresponds to backscattered light). This gives $({\bf k}_0-{\bf k})^2=k_0^2+k^2-2k_0k \cos\theta$, and the resonance condition
\begin{equation}
  k=k_0 R,
\end{equation}
where we denoted $R=1/[1+(k_0/k_C)(1-\cos\theta)]$ and $k_C=\omega_C/c$. Expressed in terms of the wavelengths
$\lambda_0=2\pi/k_0$ and $\lambda=2\pi/k$, Eq. (12) is equivalent to $\lambda-\lambda_0=(2\pi c/\omega_C)(1-\cos\theta)$, which recovers Compton's \cite{Compton} result in the limit $\gamma_A=1$.

\begin{figure}[htb]
\centering
\includegraphics[width=8.5cm]{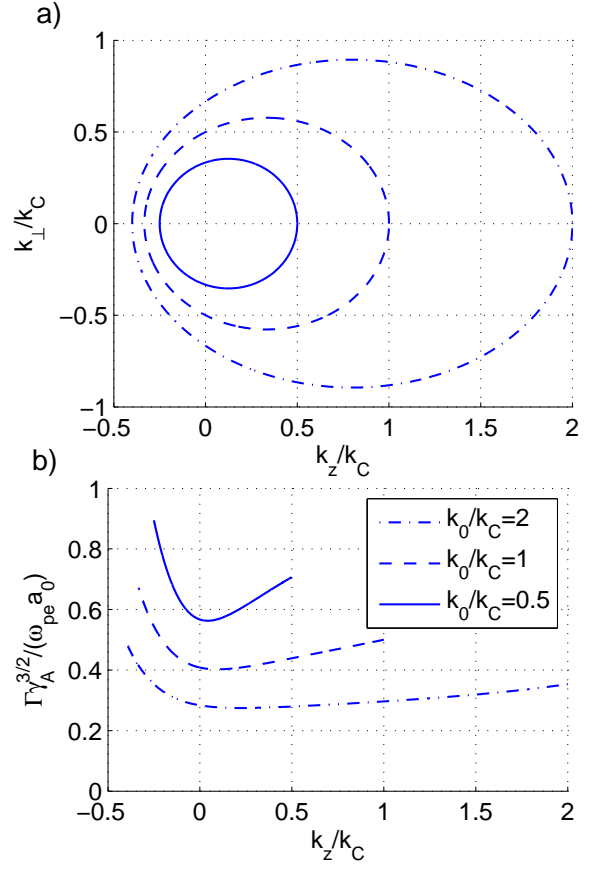}
\caption{a) The resonant radiation wave vector components $k_z$ and $k_\perp$ for $k_0/k_C=0.5$, $k_0/k_C=1$, and $k_0/k_C=2$,
where $k_C=\gamma_A m_e c/\hbar$ is the relativistic Compton wavenumber. Negative values on
$k_z$ corresponds to backscattered light. b) The normalized growth rate $\Gamma \gamma_A^{3/2}/(\omega_{pe} a_0)$ as a function of the normalized wave vector component $k_z/k_C$.}
\label{fe_Erg_0.8_eV}
\end{figure}

The resonance condition (12) can be used to express the other quantities as functions of $k_0$ and $\theta$ via the relations $\omega=ck=ck_0R$, $\widetilde{\Omega}=\omega_0-ck=ck_0(1-R)$,
$\widetilde{K}_z=k_0-k \cos\theta=k_0(1-R\cos\theta)$
and $\widetilde{K}_\perp=-k \sin\theta=-k_0 R \sin\theta$,
so that $\widetilde{K}^2=\widetilde{K}_z^2+\widetilde{K}_\perp^2=k_0^2(1+R^2-2R\cos\theta)$.
This gives $D_A(\omega,{\bf k})\approx2i\Gamma c k_0 R$, $D_L(\Omega,{\bf K})\approx-(i\Gamma c k_0(1-R)/2\omega_C^2)(4c^2 k_0^2 R(1-\cos\theta)+\omega_{pe}^2/\gamma_A+4\omega_C^2)$, and
$D_A(\Omega,{\bf K})\approx 2 c^2 k_0^2 R(1-\cos\theta)+\omega_{pe}^2/\gamma_A$.
Furthermore, for circularly polarized waves ${\bf A}_0=(\widehat{\bf x}+i\widehat{\bf y})A_0$ we evaluate $|{\bf A}_0\times {\bf k}|^2/k^2=(1+\cos^2\theta)|A_0|^2$.
Inserting these expressions into Eq. (1) gives, in the limit $\omega_{pe}^2/\gamma_A\ll c^2 k_0^2 R (1-\cos\theta)$, the growth-rate
\begin{equation}
  \frac{\Gamma}{\omega_{pe}}=\frac{(1+\cos^2\theta)^{1/2} a_0}{\sqrt{8} \gamma_A^{3/2} R^{1/2}
  (k_0/k_C)^{1/2}[(k_0/k_C)^2 R(1-\cos\theta) +1]^{1/2}}.
\end{equation}

In Fig. 1, we have plotted the components of the resonant radiation wave
vector ${\bf k}$, and the corresponding growth-rate of the instability.
The growth-rate varies slightly with the angles of the scattered light, and
has a maximum for back-scattered light where $\theta=\pi$.
As an example, we take coherent Compton radiation with a wave frequency
equal to the Compton frequency, $\omega_0=\omega_C\approx 7.8\times 10^{20}\,\mathrm{s^{-1}}$,
corresponding to $k_0/k_C=1$ in Fig. 1. For this case, we obtain from Fig. 1b the
growth rate $\Gamma\approx 0.5\omega_{pe}a_0$. As a measure of the interaction length,
we use 10 e-foldings of the instability, $L=10 c/\Gamma\approx 10 c/(0.5\omega_{pe}a_0)$.
A competing process to the collective growth-rate is
the incoherent Compton scattering of photons off individual electrons. As an estimate,
we use the Thomson cross-section $\sigma_T=(8\pi/3)r_0^2\approx 6.65\times10^{-29}\,\mathrm{m^2}$ where the classical electron radius is $r_0=e^2/(4\pi\varepsilon_0 m_e c^2)\approx 2.8\times10^{-15}\,\mathrm{m}$. The cross-section decreases for gamma rays with $\hbar\omega_0/m_e c^2>1$, and therefore the Thomson cross-section can be seen as an upper bound for the Compton scattering.
The scattering length-scale is estimated as
$L_s=1/(n_0 \sigma_T)$. The condition for collective effects to dominate the
interaction can be expressed as $L<L_s$, which, using $L=10 c/(0.5 \omega_{pe}a_0)$, gives
$a_0>10 c n_0 \sigma_T/(0.5 \omega_{pe})$. For solid density $n_0=6\times 10^{28}\,\mathrm{m}^{-3}$
giving $\omega_{pe}=1.4\times10^{16}\,\mathrm{s}^{-1}$, we have $L<L_s\approx 0.25\,\mathrm{cm}$
for $a_0>2\times10^{-6}$, corresponding to an intensity $I=\epsilon_0 \omega_0^2 m_e^2 c^3 a_0^2/e^2>6\times10^9\,\mathrm{W/cm^2}$. Such intensities could potentially be
achieved in forthcoming experiments, and much higher intensities
exist in astrophysical settings such as gamma ray bursts etc.

Finally, we note that the solution (10) can be interpreted as an extension of
classical plasma oscillations into the
relativistic quantum regime. A pure quantum case is the solution $\widetilde{\Omega}=+\omega_C+\sqrt{\omega_C^2+c^2 \widetilde{K}^2}$, which is a high-frequency pair branch \cite{Kowalenko} with $\Omega>2\omega_C$. It could potentially lead to a 3-wave coupling for $k_0>k_C/(1+\cos\theta)$ by replacing Eq. (9) by $\omega=-ck$. However, repeating the above calculations for this case does not predict an instability, and hence the scattering of EM waves off pair modes seems not to be energetically favored in the Raman-like process discussed here.



\begin{thebibliography}{99}
\bibitem{Glenzer} {\sc Glenzer S. H., Landen O. L., Neumayer P.} {\it et al.}, {\em Phys. Rev. Lett.}, {\bf 98} (2007) 065002.
\bibitem{Glenzer09} {\sc Glenzer S. H.} and {\sc Redmer R.}, {\em Rev. Mod. Phys.}, {\bf 81} (2009) 1625.
\bibitem{Neumeyer10} {\sc Neumayer P., Fortmann C., D\"oppner T.} {\it et al.}, {\em Phys. Rev. Lett.}, {\bf 105} (2010) 075003.
\bibitem{Hand09} {\sc Hand E.}, {\em Nature (London)}, {\bf 461} (2009) 708.
\bibitem{Tanaka12} {\sc Ishikawa T., Aoyagi H., Asaka T.} et al., {\em Nature Photonics}, {\bf 6} (2012) 540.
\bibitem{Schroeder01} {\sc Schroeder C. B., Pellegrini C.} and {\sc Chen P.}, {\em Phys. Rev. E}, {\bf 64} (2001) 056502.
\bibitem{Knoll00} {\sc Knoll G. F.}, {\em Radiation Detection and Measurement} (John Wiley \& Sons, New York, 2000).
\bibitem{Baldwin97} {\sc Baldwin G. C.} and {\sc Solem J. C.}, {\em Rev. Mod. Phys.}, {\bf 69} (1997) 1085.
\bibitem{Rivlin07} {\sc Rivlin L. A.}, {\em Quantum Electron.}, {\bf 37} (2007) 723.
\bibitem{Rivlin10} {\sc Rivlin L. A.} and {\sc Zadernovsky A. A.}, {\em Laser Phys.}, {\bf 20}(5) (2010) 971.
\bibitem{Tkalya11} {\sc Tkalya E. V.}, {\em Phys. Rev. Lett.}, {\bf 106} (2011) 162501.
\bibitem{Son12} {\sc Son S.} and {\sc Moon S. J.}, {\em Phys. Plasmas}, {\bf 19} (2012) 063102.
\bibitem{Drake74} {\sc Drake J. F., Kaw P. K., Lee Y. C., Schmidt G., Liu C. S.} and {\sc Rosenbluth M. N.}, {\em Phys. Fluids}, {\bf 17} (1974) 778.
\bibitem{Shukla86} {\sc Shukla P. K., Rao N. N., Yu M. Y.} and {\sc Tsintsadze N. L.}, {\em Phys. Rep.}, {\bf 138} (1986) 1.
\bibitem{McKinstrie92} {\sc McKinstrie C. J.} and {\sc Bingham R.}, {\em Phys. Fluids B}, {\bf 4} (1992) 2626.
\bibitem{Sakharov94} {\sc Sakharov A. S.} and {\sc Kirsanov V. I.}, {\em Phys. Rev. E}, {\bf 49} (1994) 3274.
\bibitem{Guerin95} {\sc Guerin S., Laval G., Mora P., Adam J. C., Heron A.} and {\sc Bendib A.}, {\em Phys. Plasmas}, {\bf 2} (1995) 2807.
\bibitem{Adam00} {\sc Adam J. C., Heron A., Laval G.} and {\sc Mora P.}, {\em Phys. Rev. Lett.}, {\bf 84} (2000) 3598.
\bibitem{Takabayasi53} {\sc Takabayasi T.}, {\em Prog. Theor. Phys.}, {\bf 9} (1953) 187.
\bibitem{Takabayasi56} {\sc Takabayasi T.}, {\em Phys. Rev.}, {\bf 102} (1956) 297.
\bibitem{Takabayasi57} {\sc Takabayasi T.}, {\em Prog. Theor. Phys. Suppl.}, {\bf 4} (1957) 2.
\bibitem{Kuzelev11} {\sc Kuzelev M. V.}, {\em JETP}, {\bf 112} (2011) 333.
\bibitem{Eliasson11a} {\sc Eliasson B.} and {\sc Shukla P. K.}, {\em Phys. Rev. E}, {\bf 83} (2011) 046407.
\bibitem{Eliasson11b} {\sc Eliasson B.} and {\sc Shukla P. K.}, {\em Phys. Rev. E}, {\bf 85} (2012) 065401(R).
\bibitem{Kowalenko} {\sc Kowalenko V., Frankel N. E.} and {\sc Hines K. C.}, {\em Phys. Rep.}, {\bf 126} (1985) 109.
\bibitem{Compton} {\sc Compton A. H.}, {\em Phys. Rev.}, {\bf 21} (1923) 483.
\end{thebibliography}
\end{document}